\documentclass[aps,prb,reprint]{revtex4-1}
\usepackage[utf8]{inputenc}

\usepackage[hidelinks]{hyperref}
\usepackage{graphicx}
\usepackage{dcolumn}
\usepackage{bm}
\usepackage{color}
\usepackage{amsmath, amssymb}

\begin{document}

\title{Shear banding in metallic glasses described by alignments of Eshelby quadrupoles}

\author{Vitalij Hieronymus-Schmidt}
\email{vitalij.schmidt@uni-muenster.de}
\affiliation{Institut für Materialphysik, Westfälische Wilhelms-Universität  Münster, Wilhelm-Klemm-Str. 10, 48149 Münster, Germany}
\author{Harald Rösner}
\email{rosner@uni-muenster.de}
\affiliation{Institut für Materialphysik, Westfälische Wilhelms-Universität  Münster, Wilhelm-Klemm-Str. 10, 48149 Münster, Germany}
\author{Alessio Zaccone}
\affiliation{Department of Chemical Engineering and Biotechnology, University of Cambridge, New Museums Site, Pembroke Street, CB2 3RA Cambridge, U.K.}
\affiliation{Cavendish Laboratory, University of Cambridge, JJ Thomson Avenue, CB3 9HE Cambridge, U.K.}
\author{Gerhard Wilde}
\affiliation{Institut für Materialphysik, Westfälische Wilhelms-Universität  Münster, Wilhelm-Klemm-Str. 10, 48149 Münster, Germany}
\affiliation{Institute of Nanochemistry and Nanobiology, School of Environmental and Chemical Engineering, Shanghai University, Shanghai 200444, P. R. China}

\begin{abstract}
Plastic deformation of metallic glasses performed well below the glass transition temperature leads to the formation of shear bands as a result of shear localization. It is believed that shear banding originates from individual stress concentrators having quadrupolar symmetry. To elucidate the underlying mechanisms of shear band formation, microstructural investigations were carried out on sheared zones using transmission electron microscopy. Here we show evidence of a characteristic signature present in shear bands manifested in the form of sinusoidal density variations. We present an analytical solution for the observed post-deformation state derived from continuum mechanics using an alignment of quadrupolar stress field perturbations for the plastic events. Since we observe qualitatively similar features for three different types of metallic glasses that span the entire range of characteristic properties of metallic glasses, we conclude that the reported deformation behavior is generic for all metallic glasses, and thus has far-reaching consequences for the deformation behavior of amorphous solids in general.

\end{abstract}


\maketitle

Crystals have the ability to deform at constant volume along slip planes via dislocations since the periodicity of the lattice provides identical atomic positions for the sheared material~\cite{Taylor1934}. However, the situation is different for amorphous materials such as metallic glasses because of their inherent structural heterogeneity as well as the absence of topologically well-defined structural ``defects''. As a consequence, extra volume needs to form in order to accommodate the mismatch between sheared zones (shear bands) and surrounding matrix~\cite{Spaepen2006}. Such zones are softer than the surrounding matrix enabling the material to flow. Although less clearly defined on the topological and atomic-level, the free volume in amorphous materials may be thought of as a carrier of plasticity equivalent to dislocations in crystalline materials. It is commonly accepted that shear bands are associated with a structural change such as local dilatation caused by shear localization, implying a volume change and thus a change in the atomic density, $\rho$~\cite{Spaepen1977, Argon1979, Donovan1981, Li2002a, Klaumunzer2011, Greer2013, Li2013, Maass2015, Hufnagel2016}. An important issue is, therefore, the local quantification of free volume or density inside shear bands. Recently, the local density within shear bands of an Al$_{88}$Y$_{7}$Fe$_{5}$ metallic glass has been determined using high angle annular dark field scanning transmission electron microscopy (HAADF-STEM)~\cite{Roesner2014}. These experiments showed that high and low density regions alternate along the propagation direction of the shear bands with respect to the un-deformed glass matrix~\cite{Roesner2014, Schmidt2015}. Thus, densification, in addition to the expected dilatation, also occurred as a response to plastic shear deformation. So far, a theoretical and mechanistic understanding to rationalize the observed features is missing. The model presented here is capable of describing these new observations quantitatively for a series of glass forming alloys with vastly different fragility, kinetic stability and deformability in compression. Moreover, it predicts an average structural length scale of heterogeneities that control the plastic deformation of metallic glasses.

\begin{figure}
 \includegraphics[width=.45\textwidth]{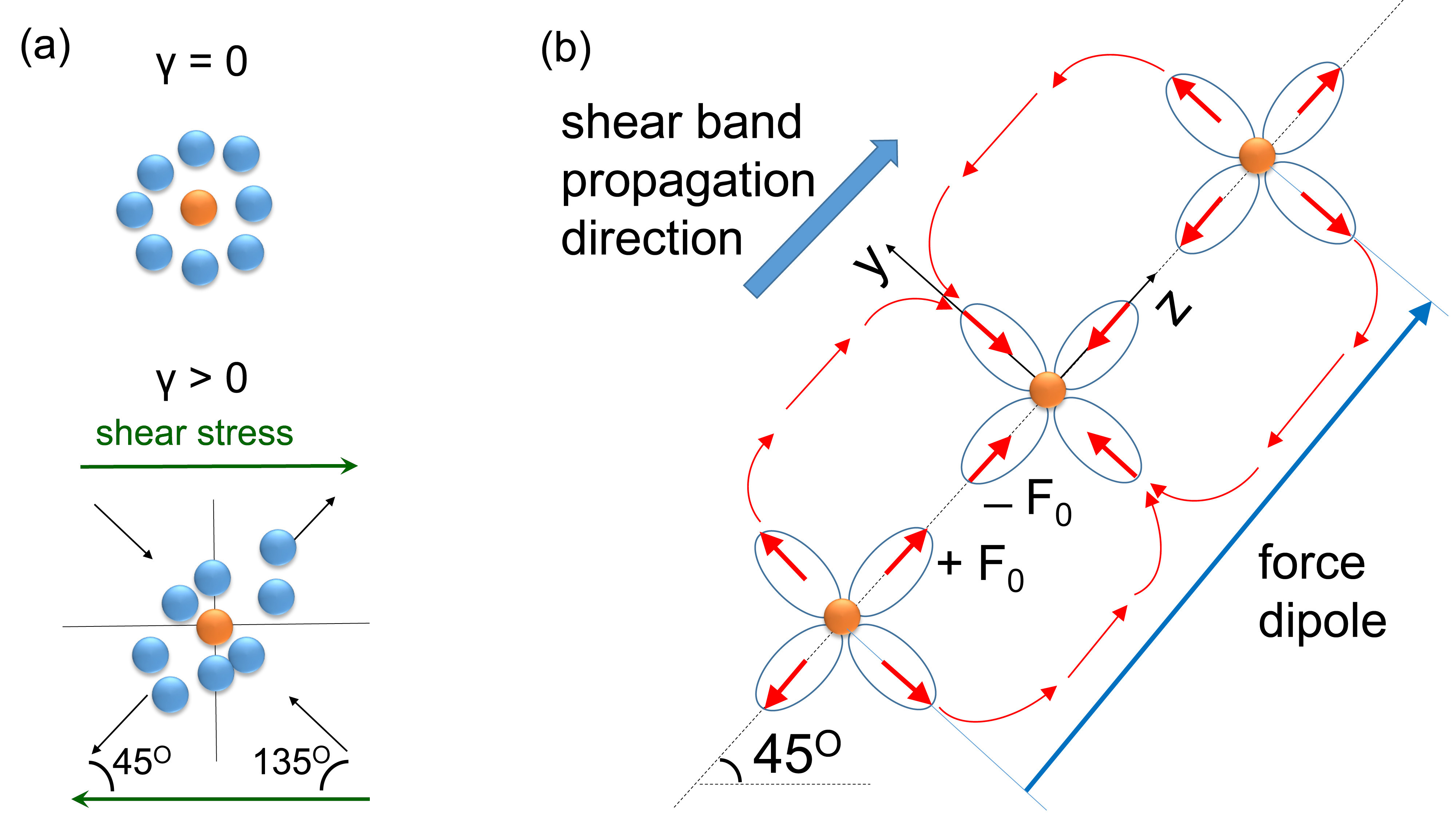}
 \caption{\textbf{(a)} Schematic of shear deformation in metallic glass. Large non-affine displacements cause particles in the shear plane to leave the glassy cage outwards along the 45$^\circ$ line or to be pushed inwards towards the centre along the 135$^\circ$ line leading to local density changes. \textbf{(b)} Illustration of the idea that density changes are caused by an alignment of Eshelby-like quadrupoles along the 45$^\circ$ direction.}
 \label{fig:1}
\end{figure}

Since the pioneering work of Spaepen and Argon~\cite{Spaepen1977, Argon1979}, it has been recognized by a large body of simulation studies that single plastic events in metallic glass are characterized by local stress fields having quadrupolar symmetry~\cite{Hufnagel2016, Maloney2006, Tsamados2008, Tsamados2010, Chikkadi2011, Jensen2014, Lemaitre2015} resembling Eshelby's famous description of inclusions within an elastic continuum~\cite{Eshelby1959}.
While the local free volume can provide the effective ``inclusions'' responsible for the Eshelby quadrupolar stress fields, one of the remaining great challenges is to understand how shear bands are actually formed in the vicinity of these stress concentrators having such complex stress fields~\cite{Packard2007}. A quadrupolar stress field symmetry is consistent with the basic symmetry of shear deformation in a disordered lattice of spherical-like atoms~\cite{Chikkadi2011}.
The inherent heterogeneity of glasses necessarily involves a distribution of atomic neighbourhoods with different effective binding strengths. Following this line of thought, the concept of local ``soft spots'' can be identified with shear-transformation zones~\cite{Spaepen1977, Argon1979, Chikkadi2011, Jensen2014, Ding2014}.
They are characterised by a weak connectivity and significant local free volume. In such a soft spot, large non-affine displacements cause particles in the shear plane to leave the glassy cage outwards along 
a particular line (in both directions), whereas they cause particles along the perpendicular line (again, in both directions) to be pushed inwards and squeezed-in towards the centre of the glassy cage (see Fig.~\ref{fig:1}a)~\cite{Zaccone2014}.

If such soft spots were aligned appropriately this would lead to alternating densities as observed in our experiments. This hypothesis is supported by recent work of Dasgupta and co-workers~\cite{Dasgupta2012}, in which MD was used to simulate an alignment of regularly spaced quadrupolar stress fields in metallic glasses. Careful examination of Fig.~4 (right-hand panel) in reference~\cite{Dasgupta2012} shows periodic density variations originating from the alignment of the quadrupoles.



\subsection{Model}
\begin{figure}[h]
 (a)\includegraphics[width=.2\textwidth]{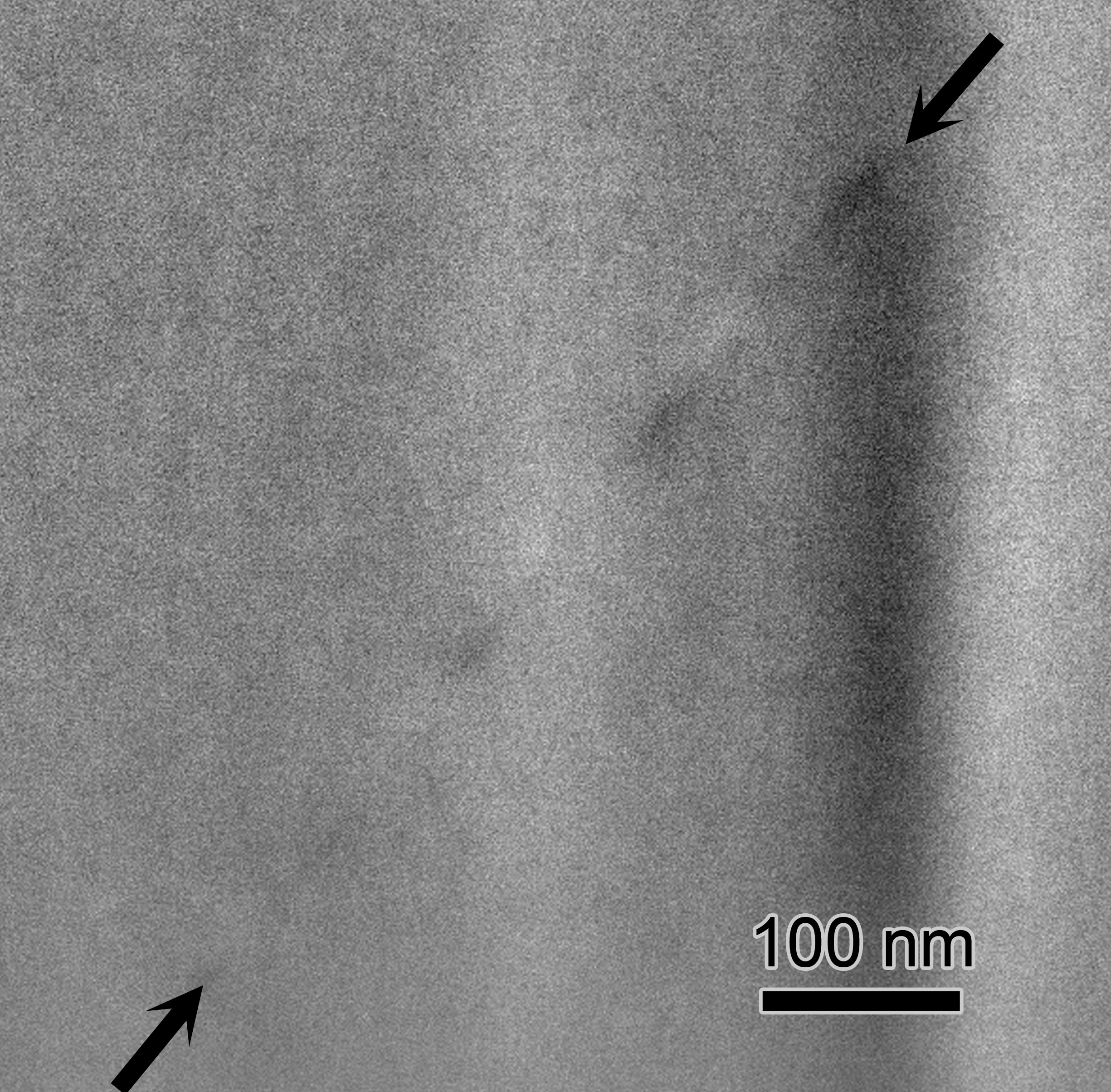} (b)\includegraphics[width=.22\textwidth]{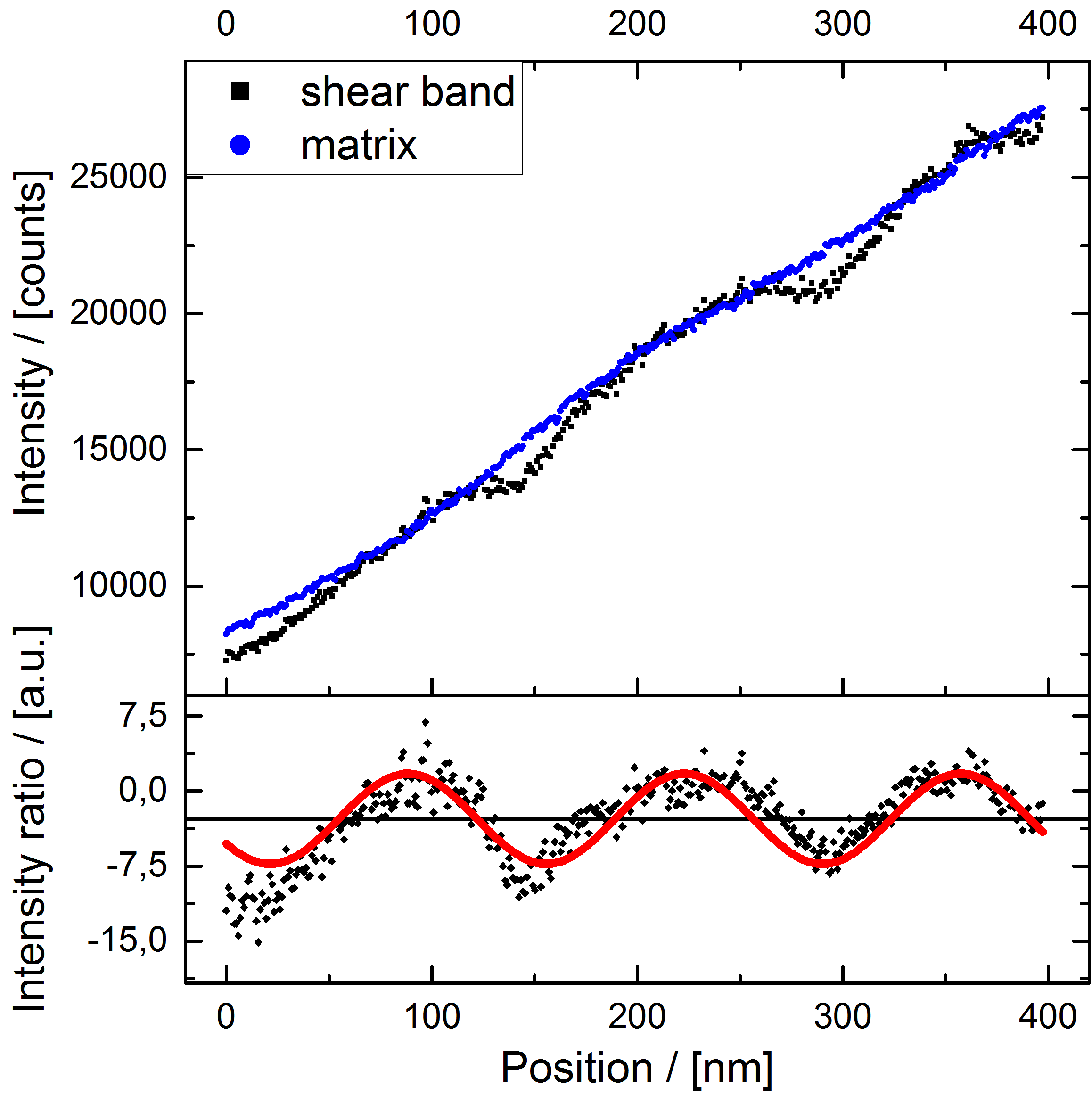}
 \caption{\textbf{(a)} HAADF-STEM image of a FIB-prepared Pd$_{40}$Ni$_{40}$P$_{20}$ bulk metallic glass sample showing contrast reversals inside a shear band (see arrows). Note that the vertical contrast (curtaining) is due to the FIB milling conditions. \textbf{(b)} Top: Corresponding intensity profiles of shear band (red) and matrix (blue). Bottom: Quantified intensity profile of the shear band relative to the matrix. The red line corresponds to the best fit using Eq.~\ref{eqn:2}.}
 \label{fig:2} 
\end{figure}

Here we propose a model based on the idea that density changes and thus shear banding are caused by an alignment of Eshelby-like quadrupoles (see Fig. 1b) that can be tested against experimental observations. We start from the basic geometry of a quadrupolar stress field perturbation for a plastic event, which locally follows a $\cos \frac{4\theta}{r^3}$ dependence, where $\theta$ is the angle, which spans the shear plane, and $r$ is the radial coordinate measured from the centre of the glassy cage.
The quadrupoles are aligned along the 45$^\circ$ directions as it has been shown analytically that such an alignment of quadrupoles minimizes the strain energy of an interacting array of Eshelby-like quadrupoles~\cite{Dasgupta2013}. It should be noted that this is an idealized situation; experiments have shown that shear band inclination angles depend on the deformation conditions (compression or tension) and can vary between 40$^\circ$ and 50$^\circ$~\cite{Ott2006}. Proceeding with our mathematical description, we label the 45$^\circ$ direction as the $z$-axis. The alignment elastic quadrupoles give rise to an alternating distribution of forces along the 45$^\circ$ direction in the shear plane (see Fig.~\ref{fig:1}b). Using
Fourier's theorem, we can write the distribution of forces $\rho_f \left( t \right)$ as a periodic function in a Fourier series expansion as $\rho_f \left( t \right) = \sum_n^\infty A_n \cos \left( k z + \varphi_n \right)$, where $A_n$ is the expansion coefficient and, $\varphi_n$ is the phase of the $n^\textrm{th}$ mode. It is worth noting while our experimentally observed density variations are not perfectly periodic, the alternations are such that we can fit them with a periodic function. The fit is better for the Pd-based glass (Fig.~\ref{fig:2}b) than for the Al-based glass (see Fig.~\ref{fig:4}a).

As a first order approximation to enable analytical calculations, we truncate the series at the first Fourier mode in the expansion. If the origin of the $z$-axis coincides with the center of the band, the phase is fixed by the symmetry to be $\varphi_n \equiv \varphi = 0$.
However, if the measurement does not start exactly at the center of the band as in the subsequent comparison with experiments, the phase will be non-zero. Hence, we approximate the force density distribution along the 45$^\circ$ $z$-axis of Fig.~\ref{fig:1} as $\rho_f \left( z \right) = A_1 \cos \left( k z \right)$, with $A \equiv A_1$, which is a normalization constant that depends on the sample size. Having defined the spatial distribution of forces, the microscopic displacement field $u \left( z \right)$ for longitudinal displacements along a given direction ($z$-axis in our case) obeys the fundamental equation of elastic equilibrium~\cite{Landau1959} $K \nabla^2 u \left( z \right) = - \rho_f \left( z \right)$, where $\rho_f \left( z \right)$ is the density of forces along $z$, which has units of force per unit volume, consistent with the left-hand side where $K$ is the bulk modulus. We now use the formal analogy between elasticity theory and electrostatics to solve for the microscopic displacement field analytically~\cite{Muggli1971}. Within this well-known analogy, as explained in detailed e.g. in Ref.~\cite{Muggli1971}, the equation for elastic equilibrium maps onto the Poisson's equation where the electrostatic potential $\phi_{el}$ is equivalent to the elastic displacement field $u$, and the bulk modulus $K$ replaces the dielectric constant $\varepsilon$.
The distribution of forces in our elastic problem plays the same role as the distribution of charges in the electrostatic problem. Hence, the distribution of forces can be written as $\rho_e \left( z \right) = \rho_f \left( z \right) = A_1 \cos \left( k z \right)$.
It can be seen that our elastic problem of microscopic displacement field along the $z$-direction in Fig.~\ref{fig:1} is mathematically identical to the problem of a 1D array of electrostatic dipoles placed along the $z$-axis. We can thus work with the Poisson's equation, formally identical to our elastic equilibrium equation, to obtain the electrostatic potential, which yields the form of the displacement field $u \left( z \right)$ of the elastic problem.
Focusing on a 1D array of electrostatic dipoles placed along the $z$-axis of Fig.~\ref{fig:1} as a consequence of the quadrupoles alignment, the charge distribution must obey the Poisson's equation which to a good approximation may be written as $\nabla^2 \phi_{el} \approx \frac{d^2 \phi_{el}}{d z^2} = \frac{\rho_f\left( z \right)}{\varepsilon}$. The qualitative form of the solution for the electrostatic field is readily obtained by inspection, and gives $\phi_{el} \sim \frac{1}{\varepsilon} \sin \left( k z + \varphi \right)$, upon omitting numerical pre-factors. A detailed and more exact 3D derivation, see Supplemental Material, gives the full quantitative solution as $\phi_{el} = \frac{A}{4 \sqrt{2 \pi} \varepsilon} \sin\left( k z + \varphi \right)$. Upon using this mathematical solution of the elastic problem, we obtain $u \left( z \right) = \frac{A}{4 \sqrt{2 \pi} K} \sin\left( k z + \varphi \right)$. Finally, the relative density change in the shear band along the 45$^\circ$ line is obtained as
\begin{equation}
 \frac{\rho \left( z \right)}{\rho} = \frac{A}{4 K \sqrt{2 \pi}} \sin \left( k z + \varphi \right)
 \label{eqn:1}
\end{equation}
where $\frac{\rho \left( z \right)}{\rho}$ is the normalized mass density change in the shear band relative to the average density $\rho$ of the surrounding un-deformed matrix. $\varphi$, as mentioned above, is an arbitrary phase shift (i.e. a fitting parameter in the following).

\subsection{Methods}
Ingots of Pd$_{40}$Ni$_{40}$P$_{20}$ were fabricated by ingot copper mould casting under argon atmosphere. The sizes of the as-cast ingots were 25~mm (length) x 10~mm (width) x 1~mm (height). Prior to casting, the ingots were cycled with boron oxide (B$_2$O$_3$) to purify the samples~\cite{Wilde1994}. The completely amorphous state of the cast samples was monitored by X-ray studies performed with a Siemens D5000 x-ray diffractometer using Cu K$_\alpha$ radiation and calorimetry using a differential scanning calorimeter (Perkin Elmer Diamond DSC) with a heating rate of 20~K/min. Subsequently, the ingots were deformed by cold-rolling to a thickness reduction of 10\%. Regions containing individual shear bands were prepared to electron-transparency using focused ion beam (FIB) (FEI Helios) milling. Microstructural characterization was performed using a Zeiss Libra 200FE transmission electron microscope operated at 200~kV in STEM mode and equipped with a Schottky field emitter, a HAADF detector (Fischione model 3000), an in-column ($\Omega$) energy filter and a slow scan CCD camera (Gatan US 4000). During the experiment electrons having a scattering angle greater than 65~mrad were collected by the HAADF detector (camera length of 720~mm) and a nominal spot size of 2~nm were used. For the fitting of the experimental data, which are summarized in Tab.~\ref{tab:1}, we used
\begin{equation}
 \frac{\rho \left( z \right)}{\rho} = \frac{A}{4 K \sqrt{2 \pi}} \sin \left( k z + \varphi \right) + y_0 \cong \frac{I_\textrm{SB} - I_\textrm{M}}{I_\textrm{M}}
 \label{eqn:2}
\end{equation}
where $I_\textrm{SB}$ and $I_\textrm{M}$ are the intensities extracted from the HAADF-STEM image shown in Fig.~\ref{fig:2}a, $A$ is a conversion/scaling factor, $K$ the bulk modulus of the investigated material, $k = \frac{2\pi}{L}$ the period, $\varphi$ an arbitrarily introduced phase shift (see above) and $y_0$ an offset value since the signal is not symmetric to the mathematical origin. This accounts for the different amplitudes of dilated and densified shear band regions observed in the experiment.

The Poisson’s ratios of the three metallic glasses (Al$_{88}$Y$_7$Fe$_5$ (partially crystallized cast material), Zr$_{52.5}$Cu$_{17.9}$Ni$_{14.6}$Al$_{10}$Ti$_5$ and Pd$_{40}$Ni$_{40}$P$_{20}$) were determined from ultrasonic measurements carried out with an Olympus 38DL Plus device.

\begin{table}
 \begin{tabular}{|c|c|c|c|}
  \hline
  Fit function & Al$_{88}$Y$_{7}$Fe$_{5}$ & Vitreloy105 & Pd$_{40}$Ni$_{40}$P$_{20}$ \\ \hline
  $k ~ \left[ 1 / \textrm{nm} \right]$ & 0.039 & 0.042 & 0.047 \\ \hline
  $L = \frac{2\pi}{k} \left[\textrm{nm} \right]$ & 163 & 150 & 135 \\ \hline
  Correlation length $L/2$ & 81.5 & 75 & 67.5 \\ \hline
  $y_0$ & -2.6 & -0.9 & -2.8 \\ \hline
 \end{tabular}
 \caption{Results of fitting Eq.~\ref{eqn:2} (see Methods) to the experimental observation for the three investigated glasses Al$_{88}$Y$_{7}$Fe$_{5}$, Zr$_{52.5}$Cu$_{17.9}$Ni$_{14.6}$Al$_{10}$Ti$_{5}$ (Vitreloy105) and Pd$_{40}$Ni$_{40}$P$_{20}$.}
 \label{tab:1}
\end{table}

\subsection{Results}
\begin{figure}[h]
  (a)\includegraphics[width=.2\textwidth]{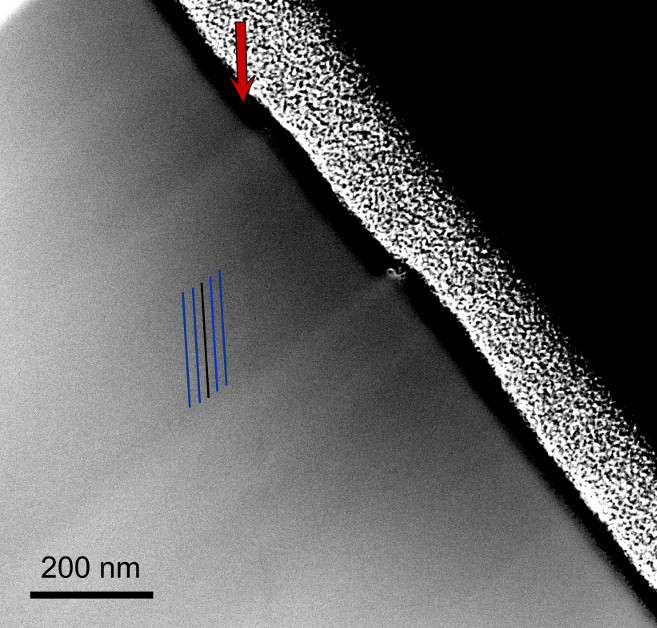} (b)\includegraphics[width=.22\textwidth]{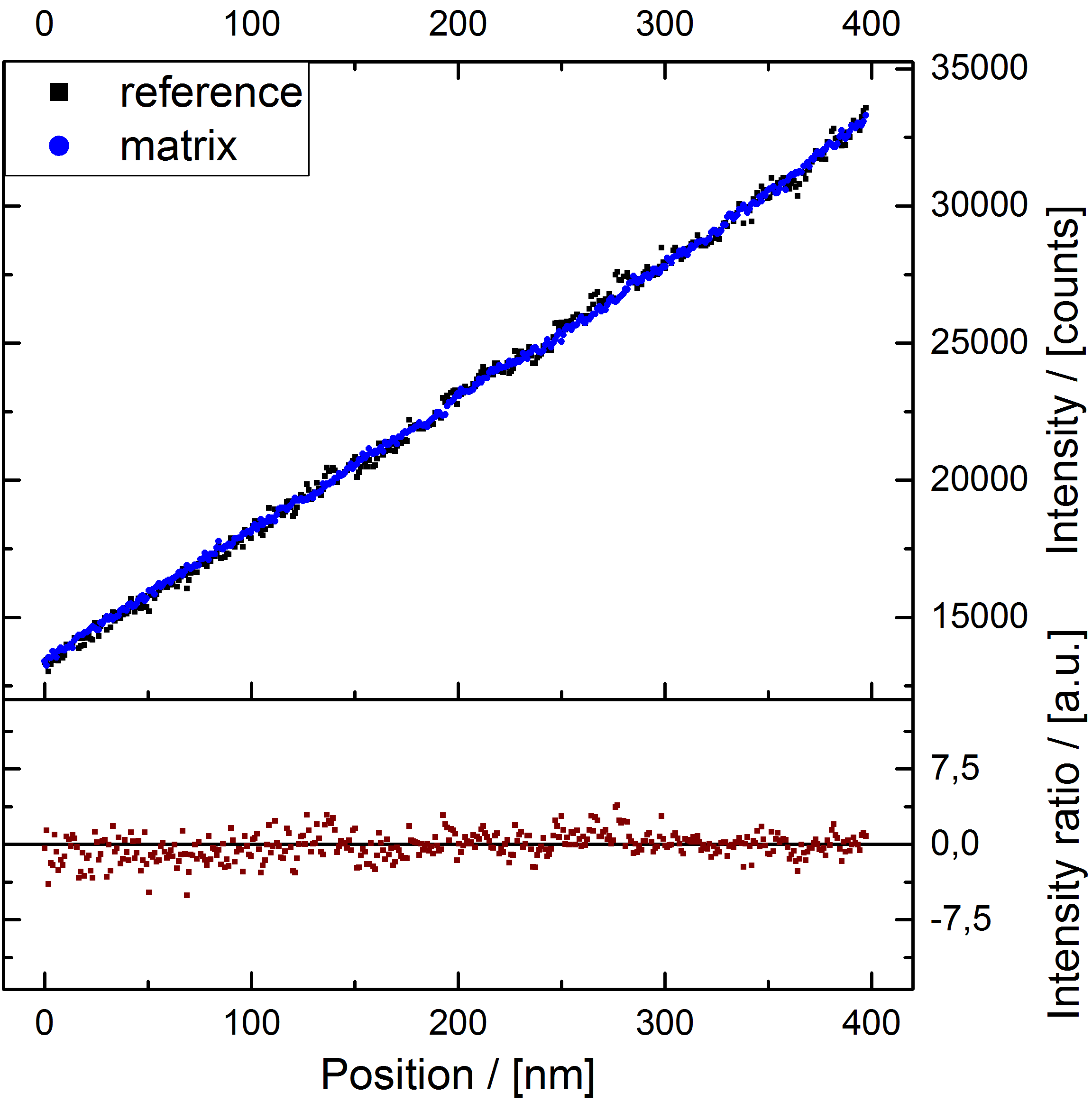}
  \caption{\textbf{(a)} HAADF-STEM overview of the FIB-prepared Pd$_{40}$Ni$_{40}$P$_{20}$ bulk metallic glass sample showing the analyzed shear band of Fig.~\ref{fig:2}a at low magnification. Note the surface off-set of the shear band (red arrow). The black and blue lines indicate the position of the reference measurement.
  \textbf{(b)} Line profiles of the reference measurement using a region next to the observed shear band. Bottom: Calculated relative intensity/density change using Eq.~\ref{eqn:2} showing scatter/noise only.}
  \label{fig:3}
\end{figure}

\begin{figure*}[!h]
  (a) \begin{minipage}{.44\textwidth}
    \includegraphics[width=\textwidth]{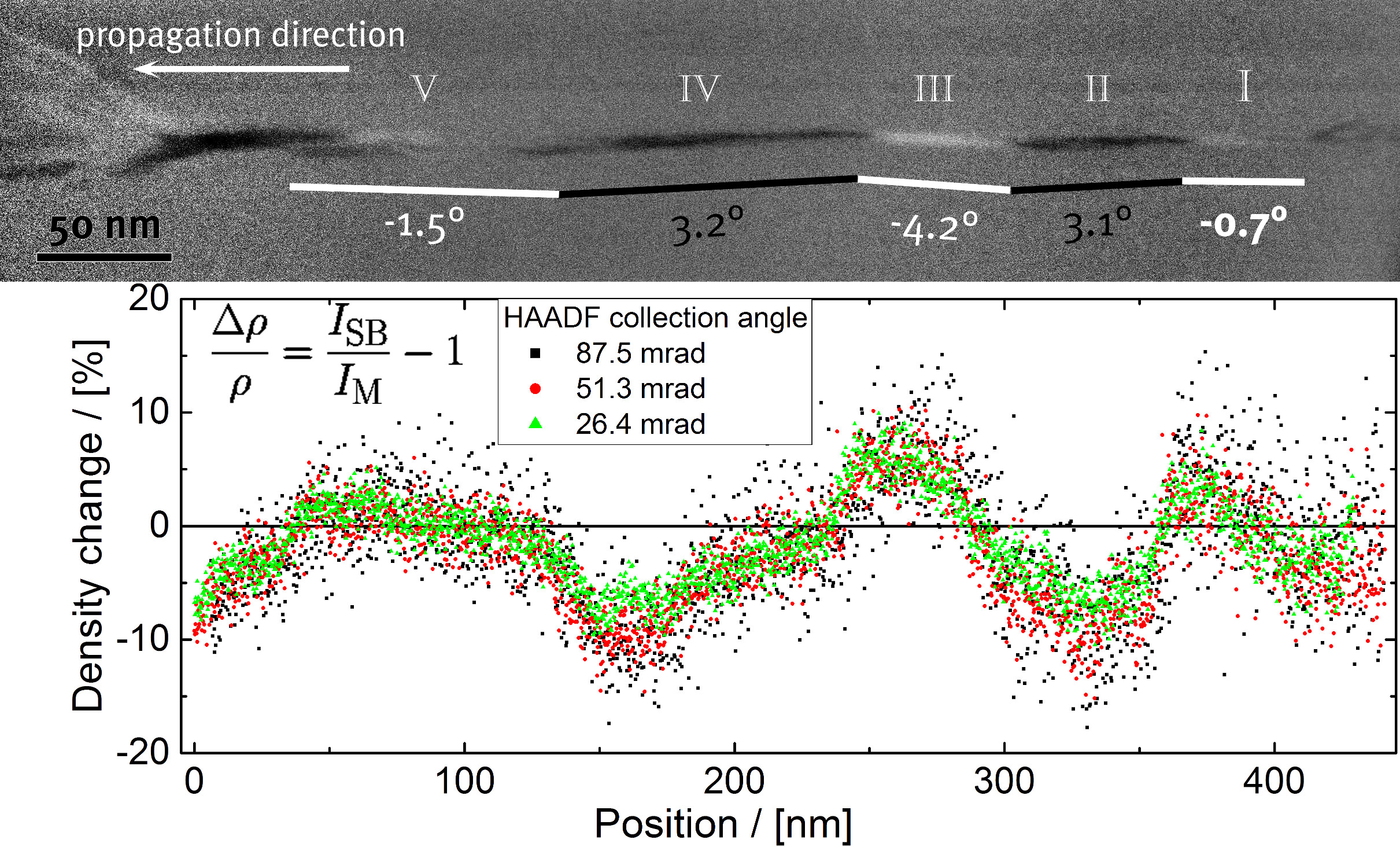}
  \end{minipage} \hspace{\fill}
  (b) \begin{minipage}{.44\textwidth}
    \includegraphics[width=.9\textwidth]{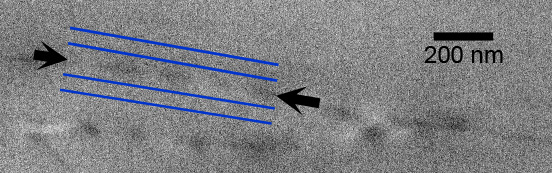} \\
    \includegraphics[width=.9\textwidth]{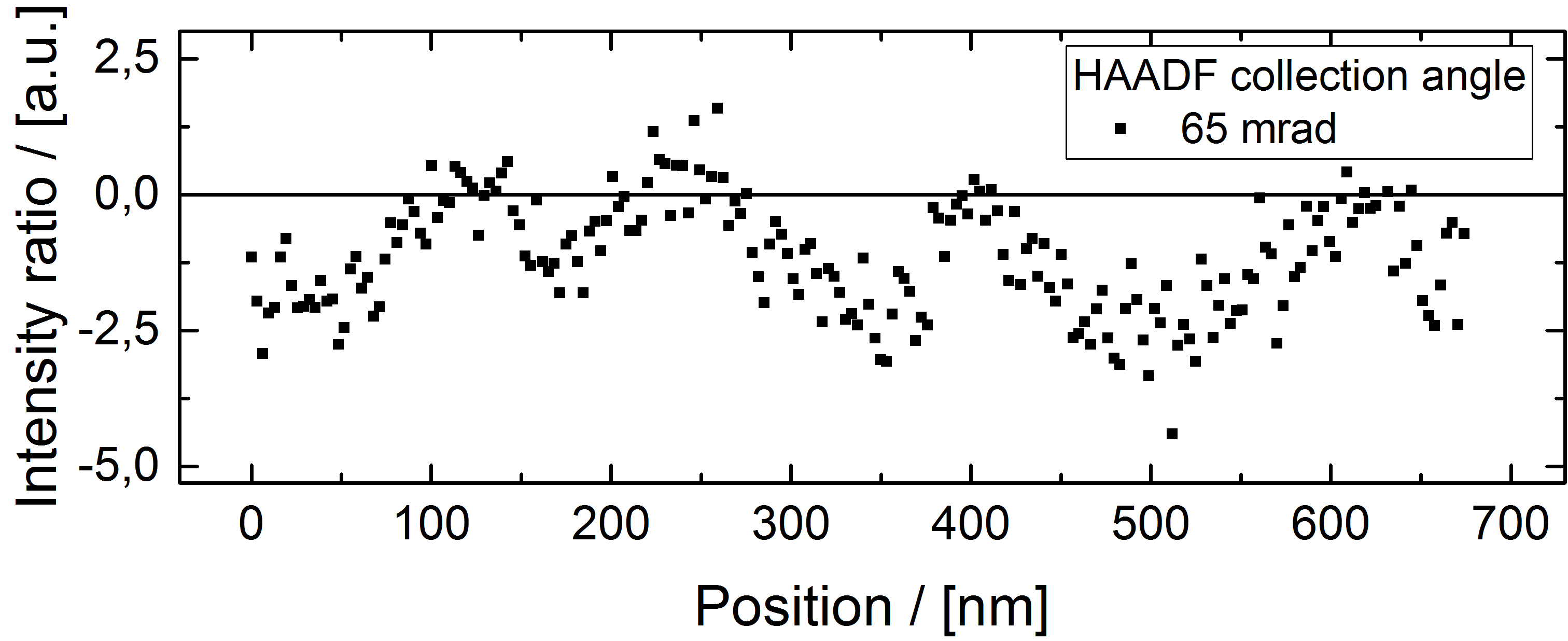}
  \end{minipage}
  \caption{\textbf{(a)} \textbf{Top}: HAADF-STEM image showing contrast reversals (bright-dark-bright) in a shear band of cold-rolled Al$_{88}$Y$_7$Fe$_5$ metallic glass. \textbf{Bottom}: Corresponding quantified density oscillations along the shear band for different collection angles of the HAADF detector. The results clearly indicate that the results are independent of the collection angle. Note that the amplitudes for the denser shear band segments are about half of the dilated states. \\ 
  \textbf{(b)} \textbf{Top}: HAADF-STEM image showing contrast reversals (bright-dark-bright) in a shear band (see arrows) of a compression-deformed bulk metallic glass sample (Zr$_{52.5}$Cu$_{17.9}$Ni$_{14.6}$Al$_{10}$Ti$_5$, Vitreloy105). \textbf{Bottom}: Corresponding quantified density oscillations along the shear band.}
  \label{fig:4}
\end{figure*}

The deformation by cold-rolling of Pd$_{40}$Ni$_{40}$P$_{20}$ bulk metallic glass produced numerous shear bands visible by the macroscopic shear off-sets at the surfaces. Individual slices of such shear bands were cut out and thinned down to electron-transparent thicknesses of about 100~nm using a focused ion beam (FIB). Fig.~\ref{fig:2}a displays part of such a FIB lamella containing a representative shear band marked by arrows and having a width of about 16~nm. The FIB lamella also displays a curtaining contrast due to the milling conditions used. The onset of the shear band was identified by the shear off-set at the surface of the foil (see Fig.~\ref{fig:3}a). It was reported that the observation of shear bands in Pd$_{40}$Ni$_{40}$P$_{20}$ is difficult~\cite{Chen2009}. While this is true, we were able to identify shear bands successfully and carefully analyse them in the following manner: A HAADF-STEM intensity profile was extracted from inside the shear band along the propagation direction as well as two on each side to determine the matrix intensity at the position of the shear band. The profile of the matrix intensity was then subtracted from the profile of the shear band intensity and the result of the difference was normalized by the profile of the matrix intensity (see Fig.~\ref{fig:2}b). (The method used for the density determination is described in more detail in the Supplemental Material and \cite{Roesner2014, Schmidt2015}). This procedure allows extraction of the density changes in the shear band relative to the matrix. For comparison, a reference measurement (see Fig.~\ref{fig:3}) following the same procedure as described above was performed at a matrix position without a shear band in order to prove that the curtaining in the TEM foil or other hidden artefacts do not cause or affect the observed periodic density variations in Fig.~\ref{fig:2}. The result of the reference measurement is shown in Supplementary Material (see Fig.~\ref{fig:3}). The result of the density variations along the real shear band is shown at the bottom of Fig.~\ref{fig:2}b. We observe small but noticeable periodic density variations with a confidence of $4\sigma$ for the smallest observed amplitude. The signal occurs periodically with larger negative and smaller positive amplitudes (Fig.~\ref{fig:2}b). Negative amplitudes correspond to dilated regions whereas the positive amplitudes refer to densified regions of the shear band compared to the surrounding matrix. Similar periodic changes between dilatation and densification (see Fig.~\ref{fig:4}) were found for a marginal glass former Al$_{88}$Y$_{7}$Fe$_{5}$ and for Zr$_{52.5}$Cu$_{17.9}$Ni$_{14.6}$Al$_{10}$Ti$_{5}$ (Vitreloy105). The periodicity of density variation is more pronounced for Pd$_{40}$Ni$_{40}$P$_{20}$ glass; however, the amplitudes of the density variations are about 10 times smaller than for the marginal glass former Al$_{88}$Y$_{7}$Fe$_{5}$. The smaller magnitude of the density changes in shear bands in some glasses seems to be the underlying reason for the difficulty in observing a distinct shear band contrast in TEM. Since we found similar observations for three very different metallic glasses, the question of an existing generic deformation mechanism for metallic glasses/amorphous solids with periodic density variations as a significant feature, is now discussed.

\subsection{Discussion}
An analytical solution of Eq.~\ref{eqn:1} using the framework of a continuum mechanics approach for the displacement field in analogy to electrostatics~\cite{Muggli1971} fits the experimental observation well since it accounts for the sinusoidal density distribution along the propagation direction of shear bands as well as for the (non-trivial) difference in amplitude for dilated and densified regions. The accordance between model and experiment strongly suggests that the density changes observed in the post deformation state are caused by an alignment of Eshelby-like quadrupoles along the shear banding propagation line (see Fig.~\ref{fig:1}b). The results are summarized in Tab.~\ref{tab:1}. We find a periodic length of about 135~nm for the Pd-based bulk metallic glass compared to about 163~nm for the marginal Al-based glass former and 150~nm for the Zr-based bulk metallic glass. 

\begin{table*}[t]
 \begin{tabular}{|l|c|c|c|c|c|c|c|c|c|}
  \hline
  & $\rho$ [g/cm$^3$] & $\nu$ & $K$ [GPa] & $G$ [GPa] & $E$ [GPa] & $T_g$ [K] & $T_x$ [K] & $T_x - T_g$ [K] & $L/2$ [nm] \\
  \hline
  Al$_{88}$Y$_{7}$Fe$_{5}$~\cite{Bokeloh2008} & 3.12\footnote{measured on partially crystallized cast material} & 0.30\footnotemark[1] & 82\footnotemark[1] & 38\footnotemark[1] & 98\footnotemark[1] & 479 & 522 & 43 & 81.5 \\
  \hline
  Zr$_{52.5}$Cu$_{17.9}$Ni$_{14.6}$Al$_{10}$Ti$_{5}$ & 6.60 & 0.37 & 113 & 33 & 91 & 681 & 743 & 62 & 75.0 \\
  \hline
  Pd$_{40}$Ni$_{40}$P$_{20}$~\cite{Nollmann2014} & 9.47 & 0.41 & 186 & 36 & 101 & 575 & 641 & 66 & 67.5\\
  \hline
 \end{tabular}
 \caption{Table of properties for the three investigated metallic glasses: The list contains the density $\rho$, Poisson’s ratio $\nu$, bulk modulus $K$, shear modulus $G$, Young’s modulus $E$, the glass transition temperature $T_g$, first crystallization on-set temperature $T_x$, width of the supercooled liquid region $T_x - T_g$ and the average correlation length $\frac{L}{2}$ determined from the observed density oscillations in shear bands.} 
 \label{tab:2}
\end{table*}

The experimentally determined values of the periodic lengths allow calculation of an average correlation length between two Eshelby-like quadrupoles (Tab.~\ref{tab:1}). The correlation length, which yields an average value of $(75 \pm 10)$~nm, is half of the periodic length. This value corresponds to the distance between the heterogeneities in these glasses which control their plastic deformation. Intuitively, a distance of the order of 75~nm may appear to be too long for a characteristic unit in the glass. Yet, it should be kept in mind, that the first initiation of non-affine transformations occur only in the most ``fertile'' or softest spots, i.e. in regions which have a local configuration that resides in the low-coordination tail of the continuous distribution of atomic packing. Moreover, independent sub-micron strain analysis using nanodot deposition~\cite{Binkowski2015} show strain profiles with a periodicity of the order of about 100~nm switching from compression to tensile strain.
The difference in amplitude for dilated and densified regions seems to be characteristic for all three metallic glass systems. Rearranging Eq.~\ref{eqn:1}, where the compressibility $\frac{1}{K}$ can be expressed in terms of density and pressure, leads to
\begin{equation}
 \frac{\rho \left( z \right)}{\rho} = \frac{A}{4 K \sqrt{2 \pi}} \sin \left( k z + \varphi \right) =  \frac{A}{4 \sqrt{2 \pi}} \frac{1}{\rho} \frac{d\rho}{dP} \sin \left( k z + \varphi \right)
 \label{eqn:3}
\end{equation}
which clearly elucidates the $\frac{1}{\rho}$ dependence of the pre-factor. One should note that while $\rho \left( z \right)$ denotes the local position-dependent density, $\rho$ denotes the overall average density of the sample, which is uniform. It is interesting to note that the trends of the average correlation lengths and the deformability of the three different metallic glasses are also paralleled in the glass forming abilities of the three glasses (see Tab. \ref{tab:2}). Moreover, the average correlation length would also account for micro-alloying effects~\cite{Lu2004, Wang2007,Wang2011, Nollmann2016} in which minor elements can drastically change the deformability. While three different glass forming systems are not sufficient to dismiss coincidence, this observation might indicate the importance of the width of the distribution of local excess volume (or coordination) for the mechanical properties and the kinetic stability of metallic glasses.

\subsection{Conclusions}
Density variations in shear bands of metallic glasses (Pd$_{40}$Ni$_{40}$P$_{20}$, Al$_{88}$Y$_{7}$Fe$_{5}$, Zr$_{52.5}$Cu$_{17.9}$Ni$_{14.6}$Al$_{10}$Ti$_{5}$) were observed along their propagation direction having periodicities between 135-163~nm with smaller positive and larger negative magnitudes. A model, using an alignment of Eshelby-like quadrupoles as input, is presented. It crucially provides the non-trivial connections between the different magnitudes for dilated and densified regions, on one hand, and the bulk modulus and sample's density, on the other. The good accordance between model and experiment strongly suggests that the observed density changes originate from aligned Eshelby-like quadrupolar stress fields. Moreover, the model predicts an average structural length scale of heterogeneities of the order of 75~nm that control the plastic deformation of metallic glasses. Since qualitatively similar features were observed for different types of metallic glasses having different compositions and vastly different characteristics, the conclusion is drawn that alternating density variations in shear bands, resulting from the alignment of Eshelby plastic events, are fundamental for the plastic deformation of all metallic glasses, and, possibly, for all amorphous materials in general.

\begin{acknowledgments}
We are grateful to Mr. M. Köhler (MPIE Düsseldorf, Germany) for help with the FIB sample preparation and to Prof. R. Maaß (University of Illinois at Urbana-Champaign, USA) for providing a compression-deformed Vitreloy105 sample. We appreciate fruitful discussions with Prof. K. Samwer (University of Göttingen, Germany) and Dr. M. Peterlechner (University of Münster, Germany). We kindly acknowledge financial support by the DFG via SPP 1594 (Topological engineering of ultra-strong glasses).
\end{acknowledgments}

\bibliography{literature}

\section*{Supplemental information}
\subsection*{Density determination using HAADF-STEM intensities}
The dark-field intensity $\frac{I}{I_0}$ contains information about the mass thickness $\rho \cdot t$ as follows~\cite{Roesner2014}:
\begin{equation*}
 \frac{I}{I_0} = \left[ 1 - \exp \left( \frac{- N_A \cdot \sigma \cdot \rho \cdot t}{A} \right) \right] = 1 - \exp\left( \frac{- \rho \cdot t }{x_k} \right)
\end{equation*}
and for small arguments:
\begin{equation}
 \frac{I}{I_0} \cong \frac{\rho \cdot t }{x_k}
 \label{eqn:4}
\end{equation}
where  is the Avogadro's number, $\sigma$ is the total scattering cross-section, $\rho$ is the density, $t$ is the foil thickness and $A$ is the atomic weight. $x_k$ is the contrast thickness, which is defined as $\frac{N_A \cdot \sigma}{A}$. For experimental data, an acquired electron energy loss (EEL) signal allows calculation of the specimen foil thickness $t$ from the low-loss spectral region~\cite{Iakoubovskii2008}. Using Eq.~\ref{eqn:4} the relative density change normalized to the intrinsic density of the material (un-deformed surrounding matrix) can be expressed:
\begin{equation}
 \frac{\Delta \rho}{\rho} = \frac{\rho_\textrm{SB}-\rho_\textrm{M}}{\rho_\textrm{M}} = \frac{I_\textrm{SB} \cdot t_\textrm{M} \cdot x_k^\textrm{SB}}{I_\textrm{M} \cdot t_\textrm{SB} \cdot x_k^\textrm{M}} - 1
 \label{eqn:5}
\end{equation}
where $\rho_\textrm{SB}$, $\rho_\textrm{M}$ are the mass densities of the shear band (SB) and the matrix, $I_\textrm{SB}$, $I_\textrm{M}$ are the HAADF intensities, $x_k^\textrm{SB}$, $x_k^\textrm{M}$ are the contrast thicknesses and $t_\textrm{SB}$, $t_\textrm{M}$ are the corresponding local foil thicknesses of SB and matrix. A constant contrast thickness $x_k$ can be assumed here over the SB area which causes the  term to cancel out in Eq.~\ref{eqn:5}~\cite{Roesner2014}. If the foil thickness $t$ is uniform or continuously increasing as for the case of a wedge-shape specimen such as in the experiment where no preferential etching is present at the SB (Fig.~\ref{fig:2}), the terms  and  cancel out. Eq.~\ref{eqn:5} simplifies then to:
\begin{equation}
 \frac{\Delta \rho}{\rho} = \frac{\rho_\textrm{SB}-\rho_\textrm{M}}{\rho_\textrm{M}} = \frac{I_\textrm{SB} - I_\textrm{M}}{I_\textrm{M}}
 \label{eqn:6}
\end{equation}
Thus the intensity ratio equals the relative density change as shown in Fig.~\ref{fig:2}.

\subsection*{Derivation of Equation 1 in the main article}
According to Fig.~\ref{fig:1} in the main article, taking one particle as the center of the frame in the shear plane, its nearest neighbors tend to move away along the extension direction $\frac{\pi}{4}$, while they are squeezed-in along the compression direction $\frac{3\pi}{4}$. Hence, the local stress field necessarily has quadrupolar symmetry, in analogy with the Eshelby inclusion quadrupole. The alignment of quadrupoles in the 45$^\circ$ direction as schematically depicted in the main article in Fig.~\ref{fig:1}b is therefore the starting point of our mathematical description.

\subsection*{Analogy between Electrostatics and Elastostatics }
When dealing with dipoles, it is most convenient to take advantage of the analogy between elasticity and electrostatics, which then includes the use of established relations for electrostatic dipoles. As is well known, the equations of elastic equilibrium and electrostatics are formally identical, provided that for each quantity in the electrostatic problem the corresponding quantity in the elastic problem is correctly defined~\cite{Muggli1971,Zangwill2012}. In our case, we are interested in determining the local displacement field since this directly relates to density fluctuations. The quantity in the electrostatics problem that is analogous to the displacement field is the electrostatic potential, with a change of~\cite{Muggli1971}: $\phi_{el} \left( \vec r \right) \Rightarrow \vec u \left( \vec r \right)$, while the corresponding quantity for the bulk modulus is the dielectric constant $K = \epsilon$.
For an array of elastic dipoles, the displacement is non-zero only along the direction of alignment of the dipoles, hence we can treat the displacement field as a scalar, $u \left( r \right) \equiv u_z \left( r \right)$.
Furthermore, the electrostatic charges, in the electrostatic problem, play the same role as the forces in the elastic case. Just as the Poisson equation relates the electrostatic potential to the charge density distribution, the same equation with changed sign relates the displacement field to the density distribution of mechanical forces in the material.

\subsection*{Distribution of forces in the shear band along the 45$^\circ$ direction}
From Fig.~\ref{fig:1}b in the main article, an alignment of Eshelby quadrupoles causes a distribution of forces along the 45$^\circ$ axis which can be described by a periodic function.
Relabeling again the 45$^\circ$ axis as the $z$-axis, we thus write the distribution of forces $\rho_f$ (equivalent to the distribution of charges $\rho_e$ in the electrostatic problem) as a periodic function in a Fourier series
\begin{equation}
 \rho_f \left( z \right) = \sum_{n=1}^\infty A_n \cos \left( k z + \varphi_n \right)
 \label{eqn:7}
\end{equation}
since any periodic function can be expanded in Fourier series. Here $A_n$ are expansion coefficients, $k = \frac{2\pi}{L}$ is the period, while $\varphi_n$ is the phase. We can consider the first-order mode in Eq.~\ref{eqn:7} as a first-order approximation to make analytical calculations.
If the origin of the $z$-axis coincides with the center of the band, the phase is fixed by the symmetry to be $\varphi_1 = 0$. However, if the measurement do not start exactly at the center of the band as in the subsequent comparison with experiments, the phase will be non-zero. Hence we use the following approximation 
\begin{equation}
 \rho_f \left( z \right) = A \cos \left( k z \right)
 \label{eqn:8}
\end{equation}
for the density distribution of forces, with $A \equiv A_1$ as a normalization constant which depends on the size of the sample.

\subsection*{Derivation of the displacement field along the 45$^\circ$ direction}
We now use the analogy with electrostatics to solve for the microscopic displacement field analytically under the assumptions presented above. We first solve for the electrostatic potential $\phi_{el}$ for a distribution of charges given by Eq.~\ref{eqn:8}, and at the end we use the relation $u \left( z \right) = - \phi \left( z \right)$ to get to the displacement field.
The field is related to the local dipole moment $\vec p \left( \vec r \right)$ of a continuous distribution of charges $\rho \left( \vec r \right)$ via the standard relation~\cite{Zangwill2012}
\begin{equation}
 \phi_{el} \left( \vec r \right) = \frac{-1}{4 \pi \epsilon} \int_L d \vec p \left( \vec r \right) \cdot \nabla \frac{1}{\left| \vec r - \vec r_0 \right|}
 \label{eqn:9}
\end{equation}
where $\vec r_0$ labels the positions of the forces, while $\vec r$ labels the generic position in space at which the field $\vec \phi$ is evaluated.
We now take a cylindrical frame where the 45$^\circ$ axis of the shear band propagation coincides with the polar axis $z$, whereas $r$ is the radial axis (oriented along the compression direction, 135$^\circ$). Clearly, $\vec p \left( \vec r_0 \right) = p \left( z \right) \widehat {\vec {z_0}}$, because the local dipole moment is oriented along the $z$ axis.
Then we express the dipole moment by introducing the force density distribution~\cite{Muggli1971}, $\vec \tau = \frac{d \vec p \left( z_0 \right)}{d z_0}$, which is also oriented along the  axis. Therefore Eq.~\ref{eqn:9} can be rewritten as
\begin{equation}
 \phi_{el} \left( \vec r \right) = \frac{-1}{4 \pi \epsilon} \int_L d z_0 \vec \tau \cdot \nabla \frac{1}{\left| \vec r - \vec r_0 \right|}
 \label{eqn:10}
\end{equation}
where $\vec r_0 = \left[ 0, 0, z_0 \right]$, in cylindrical coordinates.

For $\vec \tau = \left[ 0, 0, \tau_0 \left( z_0 \right) \right]$ only the $z$-component is non-zero. Therefore the scalar product can be easily evaluated and the integral can be written as
\begin{equation}
 \phi_{el} \left( r, z \right) = \frac{-1}{4 \pi \epsilon} \int_L d z_0 \tau_z \left( z_0 \right) \frac{d}{d z_0} \cdot \frac{1}{\sqrt{r^2 - \left( z - z_0 \right)^2}}
 \label{eqn:11}
\end{equation}
Upon evaluating the derivative we get
\begin{equation}
 \phi_{el} \left( r, z \right) = \frac{-1}{4 \pi \epsilon} \int_L d z_0 \tau_z \left( z_0 \right) \frac{\left( z -z_0 \right)}{\left[ r^2 - \left( z - z_0 \right)^2\right]^{\frac{3}{2}}}
 \label{eqn:12}
\end{equation}
Since we are interested in the displacement field along the $z$-axis (45$^\circ$ direction) at the center of the band, we take the near-field approximation~\cite{Zangwill2012}, $r \ll \left( z -z_0 \right)$, and focus on the integral 
\begin{equation}
 \phi_{el} \left( z \right) = \frac{-1}{4 \pi \epsilon} \int_L d z_0 \tau_z \left( z_0 \right) \frac{1}{\left( z - z_0 \right)^2}
 \label{eqn:13}
\end{equation}
The density of dipole moment $\tau$ along $z$ is related to the charge density distribution along $z$, via $\rho_f \left( z_0 \right) = \frac{-d}{d z_0} \tau \left( z_0 \right)$. Using Eq.~\ref{eqn:8} we thus obtain $\tau \left( z_0 \right) = -A \left(1 + \frac{\sin\left( k z \right)}{k} \right)$. Upon putting this in Eq.~\ref{eqn:11}, we get the following expression
\begin{equation}
 \phi \left( z \right) = \frac{-A}{4 \pi \epsilon} \int_L d z_0 \left( 1 + \frac{\sin \left( k z_0 \right)}{k} \right) \frac{1}{\left( z - z_0 \right)^2}
 \label{eqn:14}
\end{equation}

Upon considering an infinite medium (or at least a macroscopic size which is much larger than the atomic scale as is always the case) $\int_L d z_0 \rightarrow \int_{-\infty}^{+\infty} d z_0$, and letting the $z_0$ coordinate start at the center of the band ($\phi = 0$) for ease of notation and without loss of generality, the integral can be evaluated analytically after recognizing that it is a standard convolution integral:
\begin{equation}
 h \left( t \right) = \int_{-\infty}^{+\infty} d t' f \left( t - t' \right) g \left( t' \right)
 \label{eqn:15}
\end{equation}
with the following straightforward identifications: $z_0 = t'$, $z=t$, $g \left( t' \right) = 1 + \frac{\sin \left( k z_0 \right)}{k}$ and $f \left( t -t' \right) = \frac{1}{\left( z_0 - z \right)^2}$. As is well known~\cite{Argon1979}, convolution integrals satisfy the following property
\begin{equation}
 h \left( z \right) = \frac{1}{2 \pi} \int_{-\infty}^{+\infty} dq e^{-iqz} \hat f \left( q \right) \hat g \left( q \right)
 \label{eqn:16}
\end{equation}
where $q$ is a dummy variable which in our case has dimensions [1/length] and $\hat f \left( q \right)$ denotes the Fourier transform of the function $f \left( z \right)$, with $q = \frac{2 \pi}{z}$.
With the previous identifications we obtain
$$\hat f \left( q \right) = \frac{- \pi}{2} q~\textrm{sgn} \left( q \right)$$
\begin{equation} 
 \hat g \left( q \right) = \sqrt{2\pi} \delta \left( q \right) + \sqrt{\frac{\pi}{2}} \frac{i\delta \left( z - k \right)}{k} - \sqrt{\frac{\pi}{2}} \frac{i\delta \left( z + k \right)}{k}
 \label{eqn:17}
\end{equation}
Hence, upon taking advantage of the convolution theorem Eq.~\ref{eqn:16} and Eq.~\ref{eqn:17}, we can find the field $\phi_{el} \left( z \right)$ along the band propagation direction by simply taking the Fourier transform of the product $\hat f \left( q \right) \hat g \left( q \right)$, which gives
\begin{equation}
 \phi_{el} \left( z \right) = \frac{-A}{8 \pi \epsilon} \sqrt{\frac{\pi}{2}} e^{-i k z} \left( e^{2ikz} - 1 \right)
 \label{eqn:18}
\end{equation}
Using the standard Euler relations, this simplifies to
\begin{equation}
 \phi_{el} \left( z \right) = \frac{-A}{8 \pi \epsilon} \sin \left( k z + \varphi \right)
 \label{eqn:19}
\end{equation}
where the phase $\varphi$ is added to fit the experimental data.

This is a central result, which shows that a periodic distribution function of forces, as a consequence of the alignment of Eshelby quadrupoles, generates a sinusoidal microscopic displacement field $u \left( z \right) \Rightarrow \phi_{el} \left( z \right)$, which upon replacing $\epsilon$ with the bulk modulus $K$ reads as
\begin{equation}
 u \left( z \right) = \frac{A}{4\sqrt{2\pi}K} \sin \left( k z + \varphi \right)
 \label{eqn:20}
\end{equation}
Here the arbitrary phase $\varphi$ takes care of any arbitrary sign convention and of the fact that the experimental measurements do not necessarily start from the center of the band (which would be $z = 0$ in our treatment). 

The parameter $k$ is the same as that modulates the period of the force distribution in the Eshelby quadrupoles array in Fig.~\ref{fig:1} and Eq.~\ref{eqn:1}, and its value depends on the atomic structure and size of the building blocks. Its value therefore varies depending on the density and composition of the material. 
Finally we obtain the relative density change in the band along the 45$^\circ$ direction as
\begin{equation}
 \frac{\Delta \rho \left( z \right)}{\rho} = 1 + u \left( z \right) = \frac{A}{4 \sqrt{2 \pi} K} \sin \left( k z + \varphi \right)
 \label{eqn:21}
\end{equation}
where $\Delta \rho \left( z \right) = \rho \left( z \right) - \rho$ and $\rho$ is the average density of the material in the band.

\subsection*{Supplementary Figures}

\end{document}